\documentclass[aps,twocolumn,floatfix,superscriptaddress]{revtex4}
\usepackage{graphicx,amsmath,units,xspace,amssymb}
\usepackage{url}
\usepackage{times}
\usepackage[sort&compress]{natbib}
\begin{document}

\title{Geometric Frustration in Buckled Colloidal Monolayers}

\author{Y. Han}
\thanks{These authors contributed equally to this work.}
\affiliation{Department of Physics and Astronomy, University of Pennsylvania, Philadelphia, PA 19104, USA}
\affiliation{Department of Physics, Hong Kong University of Science and Technology, Clear Water Bay, Kowloon, Hong Kong} 
\author{Y. Shokef}
\thanks{These authors contributed equally to this work.}
\affiliation{Department of Physics and Astronomy, University of Pennsylvania, Philadelphia, PA 19104, USA}
\author{A. M. Alsayed}
\affiliation{Department of Physics and Astronomy, University of Pennsylvania, Philadelphia, PA 19104, USA}
\author{P. Yunker}
\affiliation{Department of Physics and Astronomy, University of Pennsylvania, Philadelphia, PA 19104, USA}
\author{T. C. Lubensky}
\affiliation{Department of Physics and Astronomy, University of Pennsylvania, Philadelphia, PA 19104, USA}
\author{A. G. Yodh}
\affiliation{Department of Physics and Astronomy, University of Pennsylvania, Philadelphia, PA 19104, USA}

\date{\today}

\maketitle

\textbf{Geometric frustration arises when lattice structure prevents simultaneous minimization of local interactions.  It leads to highly degenerate ground states and, subsequently, complex phases of matter such as water ice, spin ice and frustrated magnetic materials.  Here we report a simple geometrically frustrated system composed of closely packed colloidal spheres confined between parallel walls.  Diameter-tunable microgel spheres are self-assembled into a buckled triangular lattice with either up or down displacements analogous to an antiferromagnetic Ising model on a triangular lattice.  Experiment and theory reveal single-particle dynamics governed by in-plane lattice distortions that partially relieve frustration and produce ground-states with zigzagging stripes and subextensive entropy, rather than the more random configurations and extensive entropy of the antiferromagnetic Ising model.  This tunable soft matter system provides an uncharted arena in which the dynamics of frustration, thermal excitations and defects can be directly visualized.}

\begin{figure}[!t]
\centering
\includegraphics[width=\columnwidth]{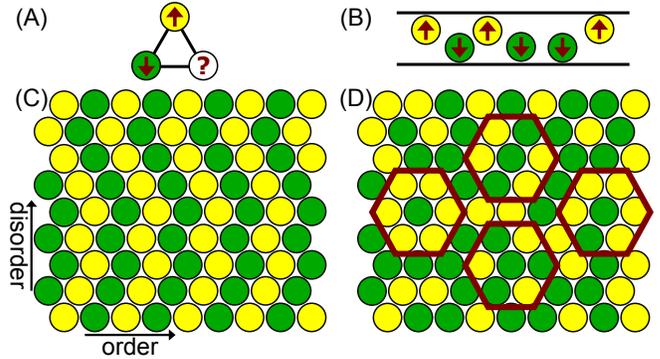}
\caption{\textbf{Ising ground state.} (A) Three spins on a
triangular plaquette cannot simultaneously satisfy all AF
interactions. (B) For colloids confined between walls separated of
order 1.5 sphere diameters (side view), particles move to opposite
walls in order to maximize free volume. (C,D) Ising ground state
configurations wherein each triangular plaquette has two satisfied
bonds and one frustrated bond. (C) Zigzag stripes generated by
stacking rows of alternating up/down particles with random sidewise
shifts; all particles have exactly 2 frustrated neighbours. (D)
Particles in disordered configurations have 0, 1, 2, or 3 frustrated
neighbours (red hexagons). } \label{fig:Ising_diag}
\end{figure}

Geometric frustration arises in physical and biological systems\cite{Moessner06} ranging from water\cite{Pauling35} and spin ice\cite{Harris97} to magnets\cite{Bramwel01,Moessner01}, ceramics\cite{Ramirez03}, and high-$T_c$ superconductors\cite{highTc}.  The essence of this phenomenon is 
best captured in the model of Ising spins arranged on a
two-dimensional (2D) triangular lattice and interacting
anti-ferromagnetically\cite{Wannier50,Houtappel}; two
of the three spins on any triangular plaquette within this lattice
can be antiparallel to minimize their anti-ferromagnetic (AF)
interaction energy, but the third spin is \textit{frustrated}
because it cannot be simultaneously antiparallel to both neighbouring
spins (Fig.~\ref{fig:Ising_diag}A).  Such frustration leads to
materials with many degenerate ground states and extensive entropy
proportional to the number of particles in the system.
Consequently, small perturbations can introduce giant 
fluctuations with peculiar dynamics.  Traditionally, these phenomena
have been explored in atomic materials by ensemble averaging
techniques such as neutron and X-ray scattering, muon spin rotation,
nuclear magnetic resonance, and heat capacity and susceptibility
measurements\cite{Moessner01}.  More recently,
artificial arrays of mesoscopic constituents have been fabricated in
order to probe geometric frustration at the single-`particle' level.
Examples of the latter include Josephson junctions\cite{Davidovic96}, superconducting rings\cite{Hilgenkamp03}, ferromagnetic islands\cite{Wang06,Moller06,Nisoli07}, and recent simulations\cite{Libal06} of charged colloids in optical traps.
Observations in these model systems, however, have been limited to
the static patterns into which these systems freeze when cooled.
Thus many questions about frustrated systems remain unexplored,
particularly those associated with single-particle dynamics.  For
example, how, when, and why do individual particles change states to
accommodate their local environments, and what kinetic mechanisms
govern transitions to glassy phases?

\begin{figure*}[!t]
\centering
\parbox{14.95cm}{\includegraphics[width=14.95cm]{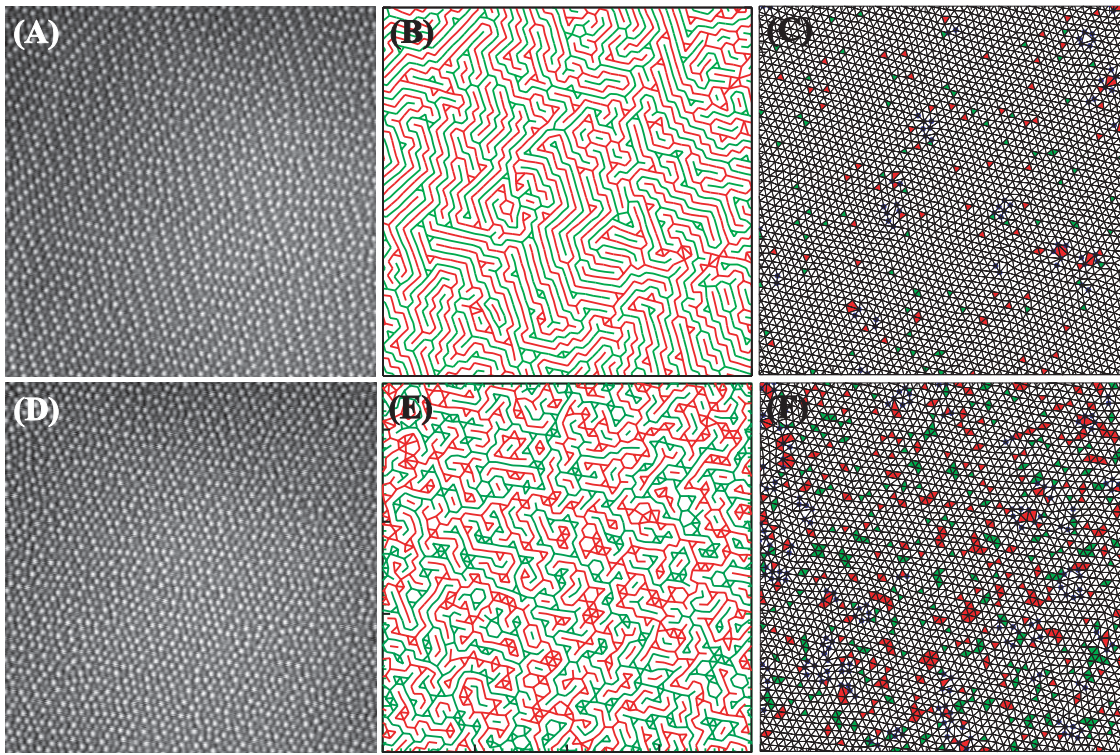}}
\parbox{2.9cm}{\caption{\textbf{Buckled monolayer of colloidal spheres. Movies are
in Supplementary Information.} $(32~\mu m)^2$ area at
$T=24.7^\circ$C (A-C) and $27.1^\circ$C (D-F). (A, D): Bright
spheres: up; dark spheres: down. (B, E): Labyrinth patterns obtained
by drawing only the frustrated up-up (red) and down-down (green)
bonds. (C, F): Corresponding Delaunay triangulations. Blue dots mark
defects in the triangular lattice, i.e. particles that do not have
exactly six nearest neighbours.  Thermally excited triangles with
three spheres up/down are labelled by red/green.}\label{fig:sample}}
\end{figure*}

Here we report on the static and \textit{dynamic} properties of a
self-assembled colloidal system analogous to Wannier's AF Ising
model\cite{Wannier50}.  Densely packed spheres between parallel
walls form an in-plane triangular lattice with out-of-plane up and
down buckling\cite{Koshikiya82,Pieranski83,Pansu84,vanWinkle86,Weiss95,Chou93,Schmidt96, Schmidt97,Zangi98,Urbach,Osterman}. The up-down states of the spheres produced by buckling are analogous to up-down
states of Ising spins (Fig.~\ref{fig:Ising_diag}B). Nearest-neighbour excluded volume interactions between particles
favour opposite states for neighbouring particles, as do the AF
interactions between neighbouring spins in the Ising model. In
contrast to engineered mesoscopic systems\cite{Davidovic96,Hilgenkamp03,Wang06,Moller06,Nisoli07}, however, the colloidal
system facilitates easy \textit{tuning} of the
effective AF interaction through changes in the diameter of
temperature-sensitive microgel spheres\cite{Alsayed05}. The colloidal system also permits direct
visualization of thermal motion at the \textit{single-particle}
level. In the limit of weak confinement, or weak
interaction strength, system properties closely follow those
predicted for the AF Ising model, but in the limit of strong
confinement, they do not. For strong
interactions, the lattice deforms to maximize free volume, and
the collective nature of the free-volume-dominated free energy
characteristic of most soft-matter systems becomes important. We
understand these effects theoretically in terms of tiling of the
plane by isosceles triangles. The tiling scheme identifies a
ground-state consisting of zigzagging stripes with sub-extensive
entropy.  Interestingly, in contrast to Ising-model predictions,
first measurements of single-particle `spin-flipping' suggest that
flipping dynamics depend not only on the number of nearest-neighbour
frustrated `bonds', but on how these bonds are arranged.  Thus the paper begins to explore connections between frustrated soft
matter and hard materials such as frustrated AF media.

\textbf{Experimental System.} For walls separated by
distances on order 1.5 sphere diameters, the particles maintain
in-plane triangular order (see Supplementary Information) but buckle out-of-plane (Fig.~\ref{fig:sample}, A, D).  This buckling minimizes system free
energy, $F=U-TS$, where $U$ is the internal energy, $T$
temperature, and $S$ entropy;  spheres move apart to lower their
repulsive interaction potential energy $U$ and to increase their
free volume $V$, which in turn leads to an entropy increase with $S
\propto \ln V$.  The effective repulsion causes spheres to move to
the top or bottom wall, and nearest-neighbours maximize free
volume by moving to opposite walls (Fig.~\ref{fig:Ising_diag}B).
Buckled colloidal monolayers were first observed more than two
decades ago\cite{Koshikiya82,Pieranski83,vanWinkle86}, and the AF
analogy was then suggested\cite{Pieranski83,Ogawa83}.
However, to date, few quantitative measurements have been performed
on this system class, and the themes explored by most of the early
work centred largely on structural transitions exhibited by
colloidal thin films as a function of increasing sample thickness\cite{Pieranski83,vanWinkle86, Weiss95,Pansu84,Schmidt96, Schmidt97,Zangi98},
rather than their connection to frustrated anti-ferromagnets.  The
use of temperature-sensitive \textit{diameter-tunable} NIPA
(N-isopropyl acrylamide) microgel spheres\cite{Alsayed05} also distinguishes our experiments from
earlier work.  By varying temperature we change particle size and
sample volume fraction and, therefore, vary the strength of the
effective AF interparticle interactions.

Samples were annealed at low volume fraction near the melting point
to produce 2D crystal domains with $\sim$$10^5$ spheres covering an
area of order (60$\mu m)^2$. Video microscopy measurements were
carried out far from grain boundaries on a $\sim$(32$\mu m)^2$
central area ($\sim$2600 spheres) within the larger crystal domain.
Particle motions were observed by microscope, recorded to
videotape using a CCD camera and tracked by
standard image-processing techniques\cite{Crocker96}.
Our colloids are very weakly charged\cite{Alsayed05}
and were measured to have \textit{short-ranged} repulsive
interactions\cite{Han07}.  Furthermore, NIPA spheres
are nearly density matched in water, so gravitational effects are
negligible.  In most colloid experiments the important thermodynamic
control variable is particle volume fraction.  The present
experiment achieved substantial variation in sphere diameter using
small changes in temperature, which altered thermal energies by less
than $1\%$.  In this paper we monitor and report temperature
rather than volume fraction because the interactions between spheres
contain a soft tail that introduces some ambiguity into the
assignment of a geometric diameter to the particles. Below 24$^\circ$C
the system is jammed and no dynamics are observed. Above
27.5$^\circ$C the in-plane crystals melt. Our primary measurements
of the frustrated states probe five temperatures from 24.7$^\circ$C
to 27.1$^\circ$C in 0.6$^\circ$C steps. In this range, the
hydrodynamic diameter of the particles decreases linearly with increasing
temperature from $0.89~\mu$m to
$0.76~\mu$m, while the average in-plane particle separation remains
$0.7~\mu$m.  To reach thermal equilibration, the sample was annealed
near the melting point before the temperature was slowly decreased.
No hysteresis was observed when slowly cycling through this
temperature range.

\textbf{Anti-Ferromagnetic Order.} The images in
Fig.~\ref{fig:sample}, A, D show roughly half of the spheres as
bright because they are in the focal plane of the microscope; the
other half, located near the bottom plate, are slightly out-of-focus
and appear dark.  We discretize the
continuous brightness profile of the particles into two `Ising' states with $s_i = \pm 1$. The
nature of the frustrated states can be exhibited in different ways
in processed images.  One way focuses on the `bonds' between particles.  We refer to
pairs of neighbouring particles in opposite
states ($s_is_j = -1$) as satisfied bonds, i.e.
satisfying the effective AF interaction, and to 
up-up or down-down pairs (with $s_i s_j = 1$) as
frustrated bonds.  Images of these bonds
show that the frustrated bonds form a nearly single-line
labyrinth (Fig.~\ref{fig:sample}B) at low temperature that then
nucleates into domains (Fig.~\ref{fig:sample}E) at high temperature.
Local AF order is alternatively characterized by the average number
of frustrated bonds per particle, $\langle N_f \rangle$. In the
limit of weak interactions, an Ising system chooses a completely
random configuration with half of the six bonds satisfied and half
frustrated, leading to $\langle N_f \rangle = 3$. In the limit of
strong interactions, on the other hand, each triangular plaquette
has one frustrated bond (Fig.~\ref{fig:Ising_diag}A),  a
third of the bonds are frustrated, and $\langle N_f
\rangle = 2$. $\langle N_f \rangle$ is a linear
rescaling of the density of excited triangles (3 up or 3 down) in
Fig.~\ref{fig:sample}, C, F, which ranges from 0 in the Ising
ground state to 0.5 for a random configuration. We find that $\langle N_f \rangle$ decreased
from approximately 2.5 to 2.1 in the temperature interval
27.1$^\circ$C-24.7$^\circ$C.  Detailed statistics of the different
local configurations are presented in Supplementary Table S1.

\begin{figure}[!t]
\centering
\includegraphics[width=\columnwidth]{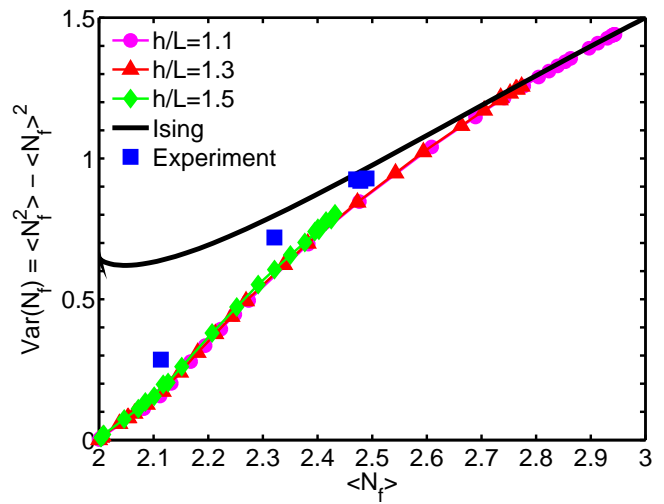}
\caption{\textbf{Fluctuation in the number of frustrated bonds per
particle as a function of its average.}  Experiments quantitatively
agree with hard-sphere simulations at different plate separations
$h$, normalized by the average in-plane lattice constant $L$.
Simulations collapse onto a single curve and deviate significantly
from the behaviour in the Ising model.} \label{fig:simulations}
\end{figure}

We first consider the static properties of the frustrated samples.
In particular we aim to identify similarities and differences
between the colloidal system and the
Ising model.  As the temperature is lowered to
increase the particle diameter, $\langle N_f \rangle$ is observed to
approach $2$.  This behaviour is expected in the Ising model ground
state.  However, the vast majority of Ising ground-state
configurations are disordered.  The colloidal monolayers, by
contrast, condense into stripe phases.  The stripes are not
straight, as could be produced by higher-order interparticle
interactions\cite{NNN}.  Rather they bend and form
zigzag patterns\cite{Schmidt96,Schmidt97,Zangi98,Urbach,Osterman} (see
Fig.~\ref{fig:sample}A and Supplementary Table S1).  In this colloidal zigzag striped phase, we
measured spatial correlations $\Gamma(i-j)=[\langle s_i s_j \rangle
- \langle s \rangle^2]/[\langle s^2 \rangle - \langle s \rangle^2]$
over separations $|i-j|$, along the principal lattice directions, of
up to 20 particles and found that they decay exponentially in
magnitude with alternating sign (Supplementary Fig.S4). 
$\Gamma(i-j)$ is positive for $i-j$ even and negative for $i-j$ odd.
In contrast, $\Gamma(i-j)$ averaged over the Ising ground state is
positive when $i-j$ is an integer multiple of $3$. Furthermore, for
zigzagging stripes each particle has exactly two frustrated
neighbours (Fig.~\ref{fig:Ising_diag}C), whereas in the fully
disordered Ising ground-state $N_f$ can be 0, 1, 2, or 3
(Fig.~\ref{fig:Ising_diag}D) and only the average $\langle N_f
\rangle$ is $2$. These observations suggest that fluctuations in
$N_f$, i.e. ${\rm Var}(N_f) = \langle N_f^2 \rangle - \langle N_f
\rangle^2$, might be a useful measure for distinguishing the zigzag
stripe phase observed here from the disordered Ising ground-state.
Figure~\ref{fig:simulations} plots the behaviour of ${\rm Var}(N_f)$
as a function of $\langle N_f \rangle$ for the Ising model and for
data obtained both from experiments and from hard-sphere Monte Carlo
(MC) simulations. Results from experiment and
simulation agree at both low and high volume
fraction and differ from those of the Ising model, especially at
high volume fraction wherein interactions are strong.

\textbf{Zigzagging Stripes.}  Ideal geometrically
frustrated systems, such as the AF Ising model, are highly
degenerate with extensive entropy at zero temperature. However, in real materials, subtle effects, for example
anisotropic interactions\cite{Houtappel}, long-range
interactions\cite{NNN}, boundary conditions \cite{Millane04} and lattice distortions\cite{Kardar,Chakraborty} relieve frustration. Our partially
ordered zigzag stripe phase at high volume fraction 
is an example of frustration relief by
lattice distortion. In the colloidal monolayer
the triangular packing is self-assembled, and (like atoms in real
solids) the particles are not forced to remain at fixed positions on
the lattice\cite{Osterman}. This deformability and
the fact that the free volume of the system is a collective function
of all particle positions breaks the mapping to simple Ising models
with pair-wise additive nearest neighbour interactions. In fact, the
positions of the colloidal particles may be thought of as comprising
a planar structure that crumples between the two confining planes.
This ``crumpling" leads to deformations of the planar triangular
lattice with satisfied bonds (projected onto the plane) on average
$3-4\%$ shorter than frustrated bonds. This difference is consistent
with the notion that each pair of neighbouring particles prefers to
be separated by the \textit{same fixed distance} in 3D, whether or
not their connecting bond is satisfied.

\begin{figure}[!t]
\centering
\includegraphics[width=\columnwidth]{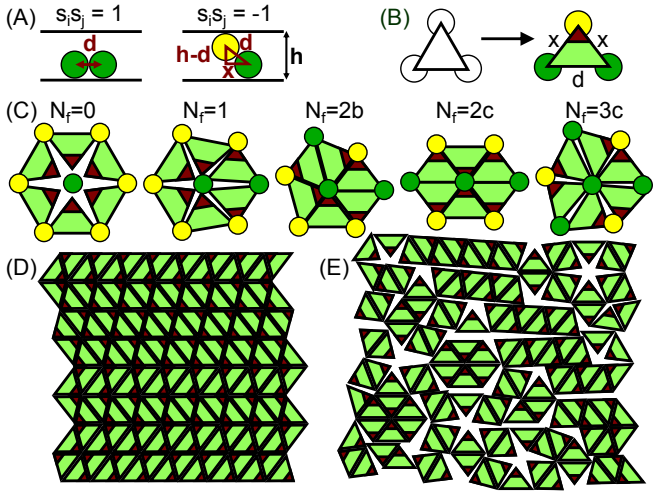}
\caption{\textbf{Tiling the plane with isosceles triangles.} (A)
Close-packed spheres are separated by one particle diameter $d$ in
3D. This distance projected on the 2D plane remains $d$ for a
frustrated bond ($s_i s_j = 1$), but is reduced to $x =\sqrt{d^2 -
(h-d)^2}$ for a satisfied bond ($s_i s_j = -1$). (B) Viewed from
above, each plaquette in the lattice tends to deform to an isosceles
triangle with one long side ($d$) along the frustrated bond and two
short sides ($x<d$) along the satisfied bonds. The angle larger than
$\pi/3$ is marked in red. (C) All possible in-plane local particle
configurations appearing in the Ising ground state. The isosceles
triangles can tile the plane without extra space only for $N_f=2$.
The ``white space'' for $N_f=0$,$1$,$3$ corresponds to additional
excluded volume.  (D, E) Tilings corresponding to striped and
disordered Ising ground-state configurations, respectively, of
Fig.~\ref{fig:Ising_diag}, C, D.} \label{fig:triangles}
\end{figure}

A simple tiling argument
demonstrates why the colloidal system ground-state configurations
of stripes and zigzags pack better than the disordered Ising
configurations. Furthermore, the tiling model shows explicitly that
maximal volume fractions of stripe and zigzag phases are the same.
Each triangular plaquette in the Ising ground-state contains two
satisfied bonds and one frustrated bond.  Thus, when spheres are
close-packed in 3D, the equilateral triangle defined by each such
triplet of neighbouring particles is tilted, and when projected onto
the 2D plane, it deforms into an isosceles triangle with two short
sides along the satisfied bonds and one long side along the
frustrated bond (Fig.~\ref{fig:triangles}, A, B).  Subsequently,
close-packed configurations of the buckled spheres in 3D are
described by tilings of the plane by isosceles triangles.
Figure~\ref{fig:triangles}C shows the configurations of isosceles
triangles for different numbers of frustrated bonds ($N_f$) in the
basic hexagonal cell. By summing up the angles around the central
vertex, one immediately sees that for $N_f=0$,$1$,$3$, the triangles
cannot close-pack. Only the two configurations with $N_f=2$ enable
tiling the plane with isosceles triangles, or, equivalently,
close-packing of the buckled spheres in 3D. Configuration 2b
corresponds to a bend in a stripe, and 2c to a stripe continuing
along a straight line. Both have the same maximal volume fraction,
thus corroborating observations of zigzagging stripes in the
experiments and simulations.

Experiments and simulations indicate a preference of the stripes to
form straight segments rather than to bend easily and thus to
generate randomly zigzagging configurations (Fig.~\ref{fig:sample}A).  Zigzagging stripes can be viewed as a
random stack of ordered lines of alternating up and down particles
(Fig.~\ref{fig:Ising_diag}C), thus straight and zigzagging
stripes are analogous to the face-centred cubic (FCC) lattice and
the random hexagonal-close-packed (RHCP) structure\cite{Pansu84} in 3D.  Straight and zigzagging stripes are
equivalent in the close-packed limit by having the same maximal volume fraction. However, for smaller volume fractions there may be an
order-by-disorder effect\cite{Moessner01,Villain80},
giving a small free volume advantage of straight stripes over
zigzagging ones, similar to the free volume advantage\cite{rhcp} of FCC over RHCP in 3D.

Instead of an extensive entropy at zero temperature\cite{Wannier50}, wherein $S$ scales linearly with
the number $N$ of particles in the system, here the entropy is
subextensive. The number of zigzagging striped configurations grows
exponentially with the linear dimension of the system
(there are two possible ways of placing one row relative to its
predecessor in Fig.~\ref{fig:Ising_diag}C), hence the entropy scales\cite{Liebmann} as $\sqrt{N}$. Alternatively put, a
non-branching single-line labyrinth is dictated by the particles on
the boundary, and for the system to rearrange from one zigzag stripe
configuration to another, a percolating cluster of order $\sqrt{N}$
particles should be flipped.

\begin{figure}[!t]
\centering
\includegraphics[width=0.9\columnwidth]{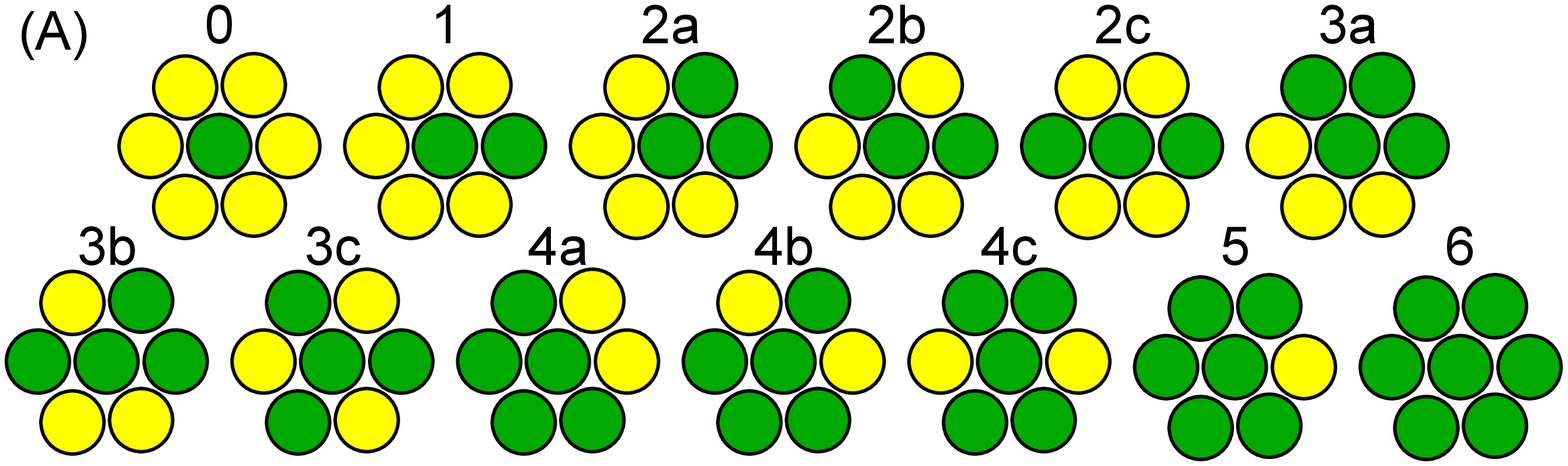}
\includegraphics[width=\columnwidth]{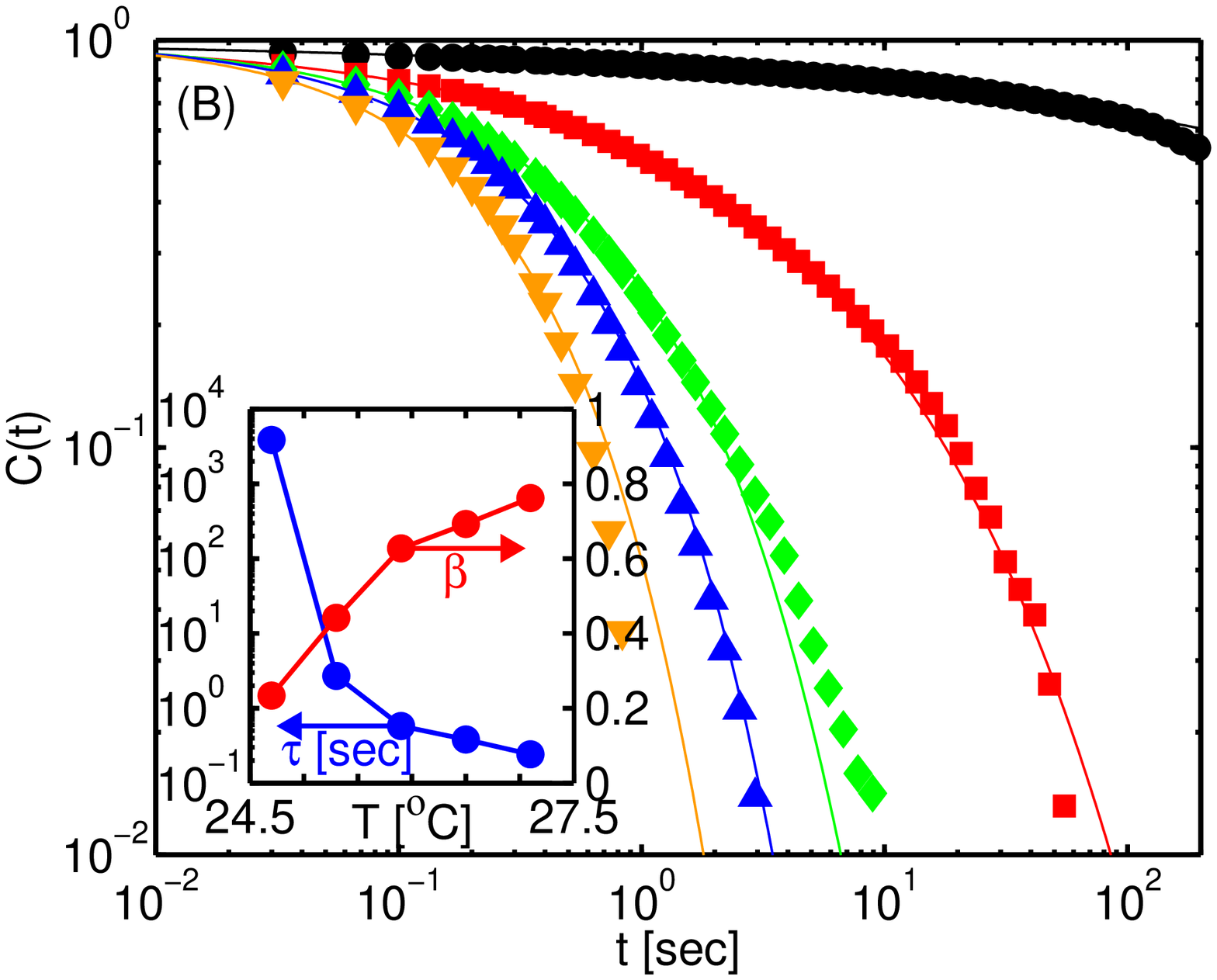}
\includegraphics[width=\columnwidth]{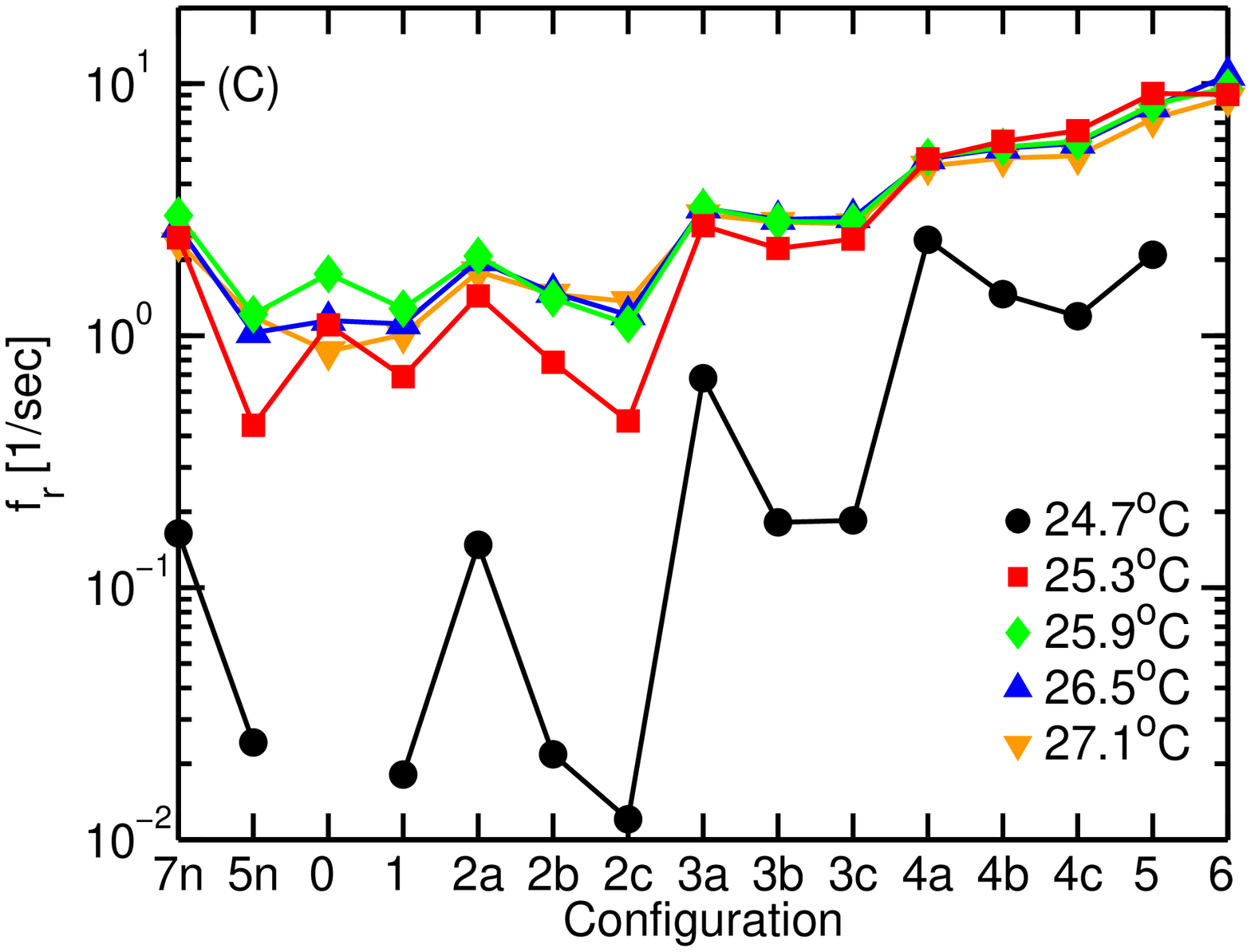}
\caption{\textbf{Single particle dynamics.} (A) Local configurations
are labelled by their value of $N_f$ and an index a,b,c indicating
the precise geometrical arrangement of the frustrated neighbours for
$N_f=2,3,4$. Symmetry to rotation and inversion reduces the $2^7$
possible configurations to the $13$ given here. (B) Single particle
autocorrelation functions plotted versus decay time. Lines are fits
to stretched exponentials $C(t) = \exp[-(t/\tau)^\beta]$, with
$\tau$ and $\beta$ given in the inset. (C) Flipping rates for the
different local environments. Configurations 7n and 5n are defects
in the in-plane lattice, with 7 and 5 nearest neighbours.}
\label{fig:dynamics}
\end{figure}

\textbf{Dynamics.}  Taken together these observations have
interesting consequences for the ground-state dynamics of frustrated
systems.  The Ising ground-state has a
local zero-energy mode, as shown in configuration 3c in
Fig.~\ref{fig:dynamics}A:  the central particle can flip without
changing the energy of the system, thus rapidly relaxing spin
correlations via a sequence of such single spin flips, even at zero
temperature.  For buckled spheres, on the other hand, the
close-packed configurations have only particles with $N_f=2$, and,
moreover, even a particle with $N_f=3$ in an excited configuration
has to cross an energy barrier in order to flip.  Like the glassy behavior of an Ising model on a deformable lattice\cite{Chakraborty00,Yin}, the slow dynamics we observe at low temperature is a consequence of the absence of local zero-energy modes in the bulk.   
Subextensive ground-state entropy also appears in related models emulating systems with glassy dynamics\cite{Nussinov04}.

Online movies (see Supplementary Information) permit us to directly
visualize `spin flipping' as well as the motions of thermal
excitations and defects in frustrated systems for the first time.
Thermal excitations labelled as coloured triangles in
Fig.~\ref{fig:sample}, C, F were typically found to be
generated/annihilated in pairs due to the flipping of a particle
shared by the two triangles. Well-isolated thermal excitations, on
the other hand, appear to be more stable. To quantify these effects,
we first extract the full time trajectory, $s_i(t)$, of each
particle $i$ from the movies. In Fig.~\ref{fig:dynamics}B we plot
the single particle autocorrelation function $C(t)= [\langle s_i(t)
s_i(0) \rangle - \langle s_i \rangle^2]/[ \langle s_i^2
\rangle-\langle s_i \rangle^2] $, averaged over all particles not at
lattice defects.  As the temperature is lowered, the correlation
function develops a stretched exponential form, $C(t) =
\exp[-(t/\tau)^\beta]$.  The measured relaxation time $\tau$
exhibits a dramatic increase as the particles swell at low
temperature, while the extracted stretching exponent $\beta$
decreases, indicating slow dynamics similar to those found in
glasses.

To further explore the dynamics of different local configurations
(defined in Fig.~\ref{fig:dynamics}A), Fig.~\ref{fig:dynamics}C shows the flipping
rate $f_r$ of single particles with a fixed neighbour structure. We measured the probability $p$
that a particle flips between consecutive images given that the
Ising states of its neighbours  remained unchanged. The time
intervals of $dt=1/30$ sec between frames were short
enough such that $p$ was typically small (0.36 at most) and the flip
rate could be approximated by $f_r = p / dt$. At high temperature,
the behaviour is similar to that of an Ising model undergoing Glauber
dynamics: $f_r \sim e^{-\Delta E/k_BT}$ where the energy difference
$\Delta E$ is proportional to the difference in $N_f$ before and
after flipping. In this regime, particles with large $N_f$
flip slightly more slowly at higher temperature because of the
weaker interactions between spheres. As the volume fraction is
increased by lowering the temperature, the particle dynamics slow by
1-2 orders of magnitude and, more interestingly,
\textit{significant differences develop between different
geometrical configurations with the same $N_f$}. Such phenomena may
not appear in the simple Ising model where the Hamiltonian depends
only on $N_f$.

Defects in the underlying lattice can strongly affect the
properties of frustrated systems. However, detailed knowledge about
the role of defects in frustrated systems is very limited. Our
experiments permit us to directly visualize defects nucleating,
annihilating, and diffusing (see Supplementary Information movies)
. By comparing trajectories containing different
numbers and types of defects, our initial studies suggest that
defect particles have enhanced in-plane diffusion 
(Supplementary Fig.S5) and slower flipping dynamics than averaging over
particles with six nearest neighbours.

\textbf{Conclusion.} We have presented experimental measurements of
single-particle dynamics in a geometrically frustrated system.
Colloidal spheres with tunable diameter self-assemble to buckled
monolayer crystals and form a system analogous to the triangular lattice AF Ising model. By tuning the volume fraction, we found that
at high compaction, in-plane lattice deformation relieves most
frustration and yields a zigzag stripe ground-state with
subextensive entropy. The `free spins' in the Ising ground state are
removed; thus the system become glassy as the volume fraction is
increased. A theoretical analysis shows that these features can
be captured by a hard-sphere model. We measured spatial
correlations and the statistics of various local configurations as
well as their flipping rates and found strong dependences on
arrangements of neighbouring particles. As the glassy phase is
approached, we observed dramatic slowing of the dynamics and
formation of stretched exponential correlation functions.
Single-defect dynamics were directly visualized and measured for the
first time. Defects have faster in-plane diffusion and slower
out-of-plane flipping than the average.

Our demonstration and analysis of this self-organized colloidal
`antiferromagnet' opens the door for the study of detailed
single-particle dynamics in frustrated systems and begins an
exploration of the connections between frustrated soft materials and
the more studied frustrated magnetic materials. Many further
manipulations can be readily applied to this system,
e.g. external control of particle motions with optical tweezers,
gravitational fields, electric fields, different particle
interactions, and defect doping. Theoretically, it will be
interesting to consider possible modifications to the Ising model
that generate a zigzagged stripe ground state and to study the
glassy dynamics arising from its subextensive zero-temperature
entropy.

\textbf{Acknowledgements} We thank Bulbul Chakraborty, Randy Kamien, Dongxu Li, Andrea Liu, Carl Modes, Tai-Kai Ng, Sturart Rice, Yehuda Snir, Tom Witten, and Yi Zhou for helpful discussions. This work is supported by NSF MRSEC grants DMR-0520020 and DMR-0505048.

\textbf{Author Information} Correspondence and requests for materials should be addressed to Y.S. (yair@sas.upenn.edu) or Y.H. (yilong@ust.hk).

\end{document}